\documentclass{article}
\usepackage{graphicx,amsmath}
\usepackage[english]{babel}

\title{
\includegraphics[width=0.35\textwidth]{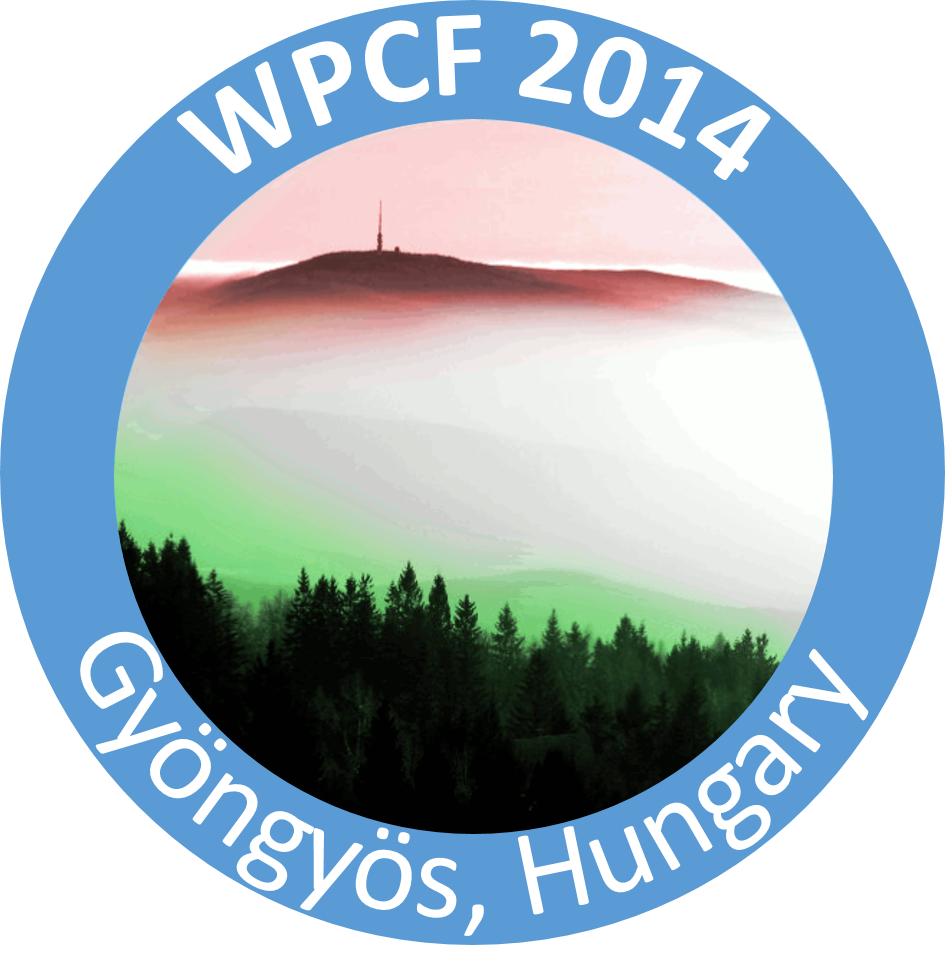}\\[1cm]
Finite size of hadrons and Bose-Einstein correlations}
\author{{Andrzej Bialas}\\[1ex]
M. Smoluchowski Institute of Physics,\\Jagellonian University, PL-30-059 Krakow\\
}

\begin{document}

\fontfamily{lmss}\selectfont
\maketitle

\begin{abstract}
In this presentation I report on the results of the paper we published recently together with Kacper Zalewski~\cite{1}. It exploits the consequences of the observation that the hadrons, being the composite objects, cannot be produced too close to each other and thus must be correlated in space-time. One of these consequences, which we discuss here, is that the correlation function need not be larger than 1 (as is necessary if the space-time correlations are absent). Since the data from LEP~\cite{2, 3} and from LHC~\cite{4} do show that the correlation function falls below 1, the particles must be correlated and we show that our observation does explain this unexpected effect. 
\end{abstract}

\begin{figure}
  \begin{center}
  \includegraphics[width=1\linewidth]{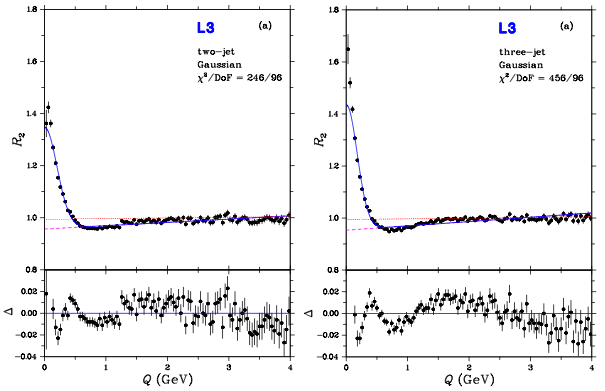}
\caption{L3 data for two-jet and three-jet events.
\label{f:1} }
  \end{center}
\end{figure}

\begin{figure}
  \begin{center}
  \includegraphics[width=0.6\linewidth]{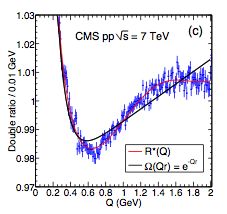}
\caption{Two-pion correlation function from CMS (pp at 7 TeV)
\label{f:2} }
  \end{center}
\end{figure}

In absence of correlations between produced hadrons, the Bose-Einstein correlation function between momenta of two identical particles 
\begin{align}
C(p_1,p_2)=\frac{N(p_1,p_2)}{N(p1)N(p2)}
\end{align}
is given by~\cite{5}
\begin{align}
C(p_1,p_2)=\frac{\tilde{w}(P_{12};Q)\tilde{w}(P_{12};-Q)}{w(p_1)w(p_2)}=
1+\frac{|\tilde{w}(P_{12};Q)|^2}{w(p_1)w(p_2)}\ge 1\label{e:2}
\end{align}
where $w(p,x)$ is the single-particle ``distribution'' (Wigner function) and 
\begin{align}
\tilde{w}(P_{12};Q)=\int dx e^{iQx} w(P_{12};x);\qquad w(p)=\int dx w(p;x),
\end{align}
$P_{12} = (p_1 + p_2)/2$; $Q = p1 - p2$.

In Fig.~\ref{f:1} the data from L3 collaboration and in Fig.~\ref{f:2} the data from CMS collaboration are shown. They clearly indicate that the correlation function $C(q)$ takes values below 1, contrary to the Eq. (\ref{e:2})

\begin{figure}
  \begin{center}
  \includegraphics[width=0.6\linewidth]{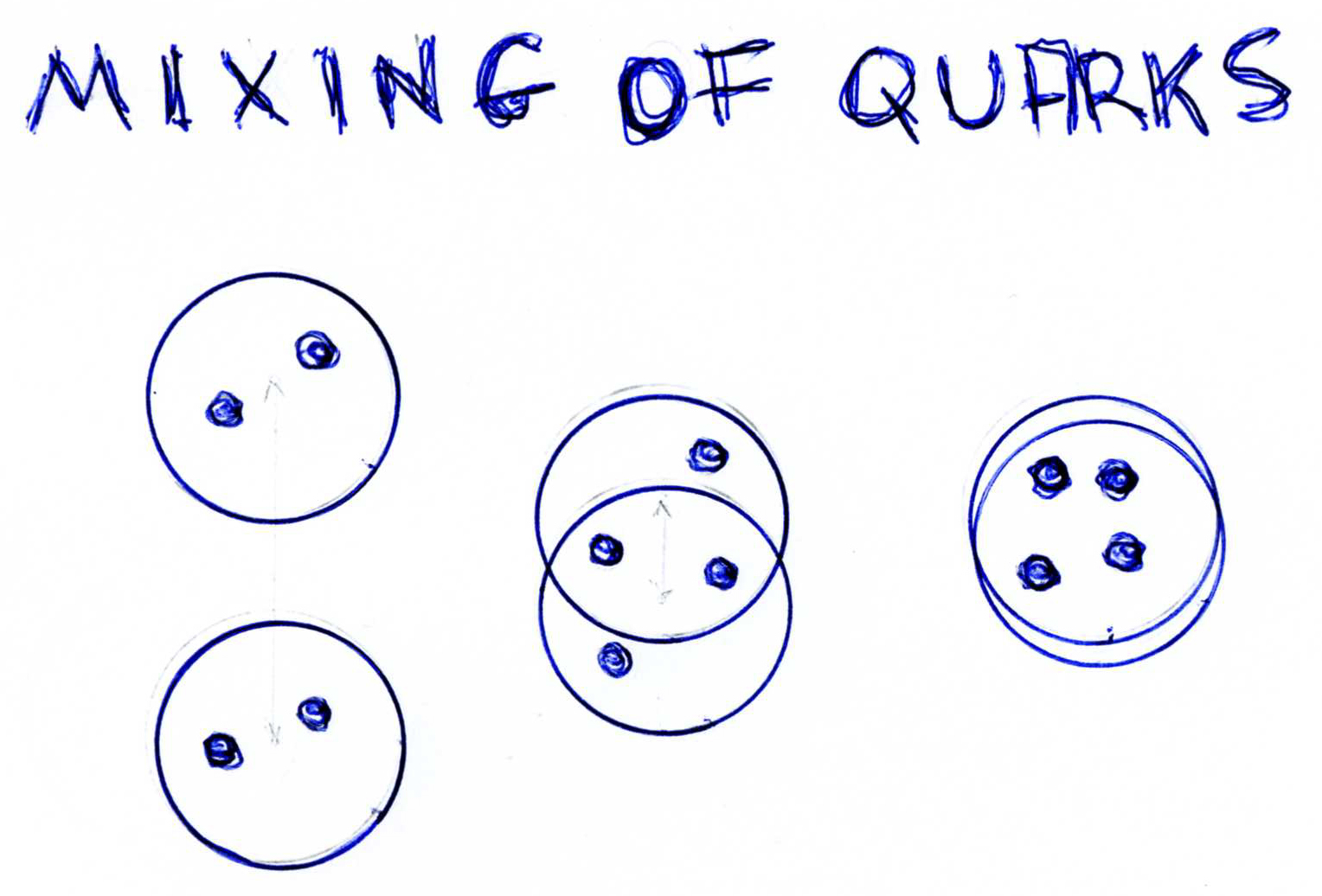}
\caption{Illustration of the excluded volume effect.
\label{f:3} }
  \end{center}
\end{figure}

These data show that particles must be correlated and we claim that the correlations responsible for this effect are caused by the composite nature of hadrons. Indeed, since hadrons are composite, they cannot be produced too close to each other because in this case \emph{they are not hadrons anymore} but rather a mixture of the hadronic constituents. This is illustrated in Fig.~\ref{f:3}. 

Since the HBT experiment measures the quantum interference between the wave functions of hadrons, it cannot see hadrons which are too close to each other. Consequently the ``source function'' $W(P12;P12;x1;x2)$ must vanish at $x_1$ close to $x_2$ and thus can be written as 
\begin{align}
w(P,x)= e^{-|\vec{x}|^2/R^2}e^{-t^2/\tau^2}f(P)\nonumber
\end{align}

\begin{align}
W(P_{12};P_{12}; x_1;x_2)= w(P_{12}; x_1)w(P_{12}; x_2)[1 - D(x_1 - x_2)]. 
\end{align}
where the cut-off function $D(x1 - x2)$ equal 1 and $(x1 - x2)$ (below, say, 1 fm) and vanishes at larger distances. 

Thus the HBT correlation function becomes: 
\begin{align}
C(p_1,p_2)=1+\frac{|\tilde{w}(P_{12};Q)|^2}{w(p_1)w(p_2)}-C_{corr}(p_1,p_2);\\
\nonumber
C_{corr}=\frac{
\int dx_1 dx_2 e^{i(x_1-x_2)Q}w(P_{12};x_1)w(P_{12};x_2)D(x_1-x_2)
}{w(p_1)w(p_2)}
\end{align}
One sees that the contribution from the part responsible for space-time correlation is negative. Moreover, since it obtains contribution from a small region of space-time, its dependence on $Q$ is much less steep than that of the uncorrelated part. Consequently, at $Q$ large enough $C(P12;Q) $may easily fall below one.

\begin{figure}
  \begin{center}
  \includegraphics[width=0.6\linewidth]{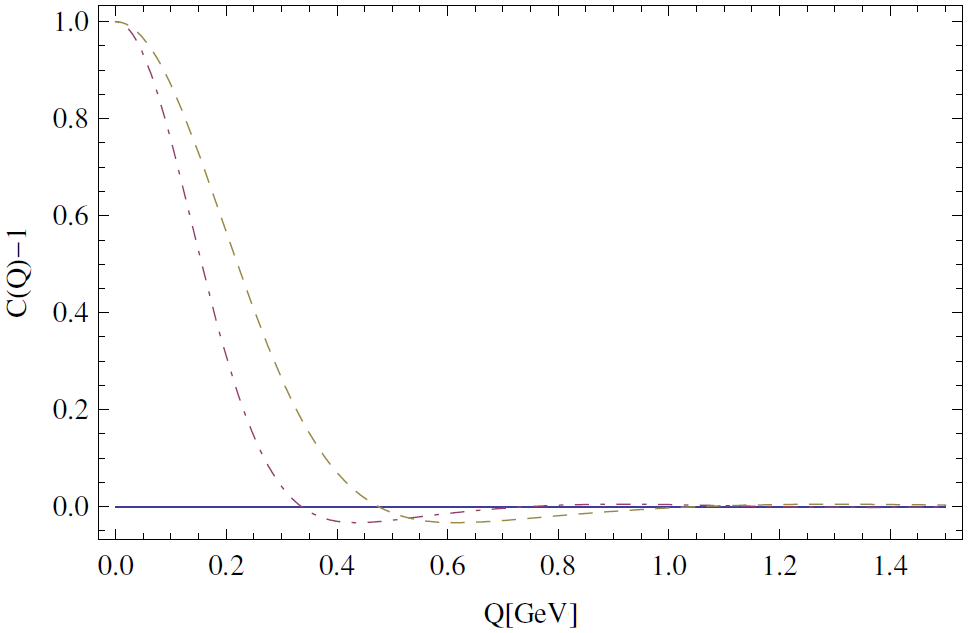}
\caption{Oscillating two-pion correlation function.
$R = r_{cut} = \tau = 1$ fm.
\label{f:4} }
  \end{center}
\end{figure}

For illustration, take $D(x_1 - x_2) =
\Theta[r_{cut}^2-|\vec{x}_1 - \vec{x}_2|^2 - (t_1 - t_2)^2]$;
The result is shown in Fig.~\ref{f:4} where one sees that, indeed, $C(Q)$ is smaller than 1 at $Q$ larger than $~400$ MeV. 

In conclusion, the presented qualitative argument shows that the observed falling of the HBT correlation function below one at large $Q$ is not accidental but re ects the fundamental fact that hadrons are NOT POINT-LIKE. Therefore this region of $Q^2$ deserves special attention in data analysis. It seems that the effect simply MUST BE THERE and the real experimental challenge is to determine its position and its size. Precise measurements may allow to determine the distance at which the hadron structure is affected by its neighbors and thus also the density at which the hadron gas starts melting into quarks and gluons.

More serious calculations, as well as a detailed comparison with data are clearly needed and are in progress (together with W.Florkowski)~\cite{6}. The preliminary results indicate that the effect significantly depends on the orientation of $Q$. This points to interest in separate measurements in \emph{side}, \emph{out} and \emph{long} directions.

\end{document}